\documentclass[sigconf, 10pt]{acmart}

\usepackage{tikz}
\usepackage{amsmath}
\usepackage{enumitem}
\usepackage{algorithm,algpseudocode}
\usepackage{filecontents}
\usepackage{booktabs} % For formal tables
\usepackage{soul}
\usepackage{graphicx}
\usepackage{caption,subcaption}
\usepackage{natbib}
  {%
      \end{minipage}%
        \end{center}}
\definecolor{mintgreen}{rgb}{0.6, 1.0, 0.6}
\definecolor{pastelviolet}{rgb}{0.8, 0.6, 0.79}
\definecolor{peridot}{rgb}{0.9, 0.89, 0.0}
\definecolor{richbrilliantlavender}{rgb}{0.95, 0.65, 1.0}
\definecolor{robineggblue}{rgb}{0.0, 0.8, 0.8}
\setcopyright{none}

\newcommand{\name}{{\em TagTeam }}
\newcommand{\names}{{\em TagTeam}}

\begin{document}
\title{\names: Towards Wearable-Assisted, Implicit Guidance for Human--Drone Teams}
% \titlenote{Produces the permission block, and
%   copyright information}
% \subtitle{Extended Abstract}
% \subtitlenote{The full version of the author's guide is available as
%   \texttt{acmart.pdf} document}

\author{Kasthuri Jayarajah\texorpdfstring{$\dag$}{[dag]}
, Aryya Gangopadhyay\texorpdfstring{$\dag$}{[dag]}
, Nicholas Waytowich\texorpdfstring{$\ddag$}{[ddag]}
}
% \authornote{Dr.~Trovato insisted his name be first.}
% \orcid{1234-5678-9012}
\affiliation{%
  \institution{\texorpdfstring{$\dag$}{[dag]}
University of Maryland, Baltimore County\\ \texorpdfstring{$\ddag$}{[ddag]}US Army Research Lab}
%   \streetaddress{P.O. Box 1212}
%   \city{Baltimore} 
%   \state{Ohio} 
\country{USA}
%   \postcode{43017-6221}
}
% \email{kasthurij,dhanujaw,archanm@smu.edu.sg}

\begin{abstract}
The availability of sensor-rich smart wearables and tiny, yet capable, unmanned vehicles such as nano quadcopters, opens up opportunities for a novel class of \emph{highly interactive}, \emph{attention-shared} human--machine teams. Reliable, lightweight, yet passive exchange of intent, data and inferences within such human--machine teams make them suitable for scenarios such as search-and-rescue with significantly improved performance in terms of speed, accuracy and semantic awareness. In this paper, we articulate a vision for such human--drone teams and key technical capabilities such teams must encompass. We present \names, an early prototype of such a team and share promising demonstration of a key capability (i.e., motion awareness).
\end{abstract}

% The default list of authors is too long for headers}
% \renewcommand{\shortauthors}{B. Trovato et al.}
\maketitle

%
% The code below should be generated by the tool at
% http://dl.acm.org/ccs.cfm
% Please copy and paste the code instead of the example below. 
%

% \keywords{ACM proceedings, \LaTeX, text tagging}

\section{Introduction}
\label{sec:intro}

\begin{figure*}[t]
% \begin{minipage}{\textwidth}
% \begin{figure}[t]
\centering
        % \begin{subfigure}[t]{0.7\textwidth}
	  \centering \includegraphics[scale=0.5]{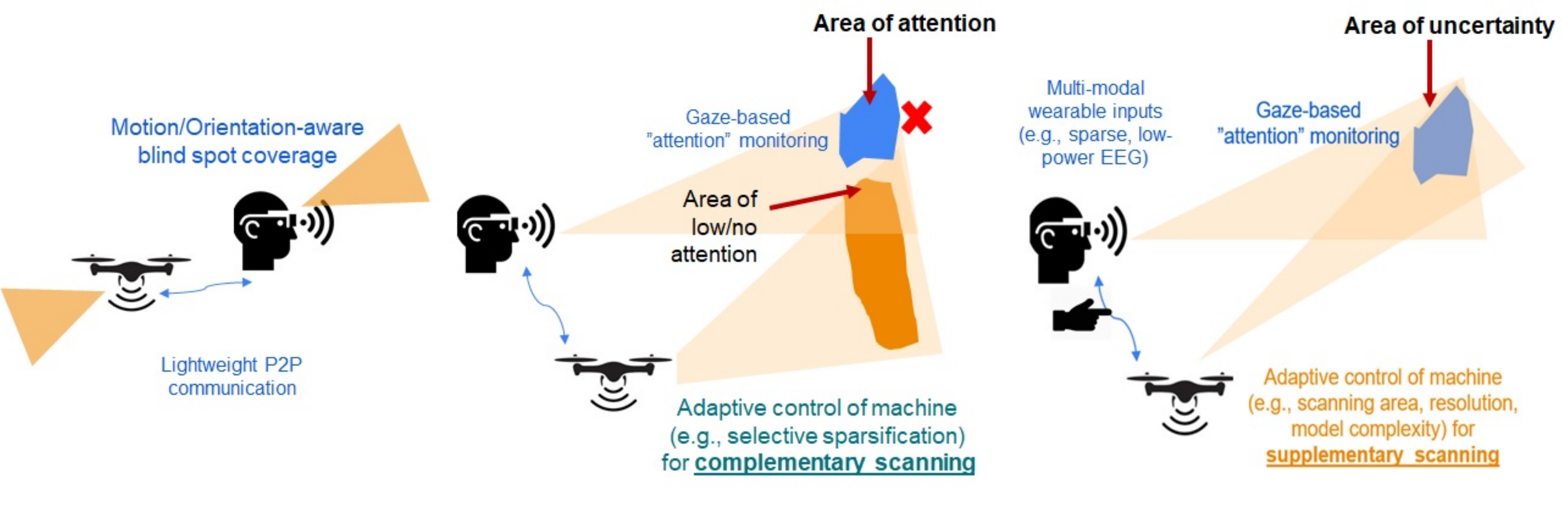}
    %   \caption{Gaze-Assisted Cooperative Visual Scanning in Indoor Environments} 
    %   %\vspace{-0.2in}
    %   \label{fig:tagteam-modes}
        % \end{subfigure}%
%         \hfill%
%         \begin{subfigure}[t]{0.3\textwidth}
% 	   \centering \includegraphics[width=\linewidth]{Figs/placeholder.png}
%     \caption{Another scenario}
% \label{fig:scenario2}
%         \end{subfigure}%
%  \vspace{-0.1in}
% \label{fig:yyyy}
% \end{figure}
\caption{Illustrative \name Scenarios: Gaze-Assisted Cooperative Visual Scanning in Indoor Environments. See \hl{\hbox{https://youtu.be/KYeo2Aichgs}} for a video demonstration of the blind spot coverage scenario.}
    \vspace{-0.1in}
\label{fig:tagteam-modes}
% \end{minipage}
\end{figure*}

The availability of sensor-rich smart wearables and tiny, yet capable, unmanned vehicles such as nano quadcopters\footnote{https://www.bitcraze.io/}, opens up opportunities for a novel class of \emph{highly interactive}, \emph{attention-shared} human--machine teams. Reliable, lightweight, yet passive exchange of intent, data and inferences within such human--machine teams make them suitable for scenarios such as search-and-rescue with significantly improved performance in terms of speed, accuracy and semantic awareness. 

Our proposed paradigm of human--drone teams are motivated by two salient trends:
\begin{itemize}[leftmargin=*]
\item \emph{\textbf{Sensor-Rich, Pervasive Wearable Devices: }} Whilst smart watches, bands and rings have been widely studied for fine-grained, gesture-based control of smart environments and machines, more recent technologies are capable of continuously capturing more than just inertial motion. For instance, miniaturized sensors such as earables (i.e., wearables worn on the ear such as the Emotiv MN8\footnote{https://www.emotiv.com/setup/mn8/}) and smart glasses (e.g., AttentiveU\footnote{https://www.media.mit.edu/projects/attentivu/overview/}) are equipped with a range of inertial and physiological sensors that can measure brain activity, eye movements, etc. that can passively measure the wearer’s cognitive processes. More recently, devices such as the Microsoft HoloLens 2\footnote{https://docs.microsoft.com/en-us/hololens/hololens2-hardware} embedded with a variety of vision, depth and time-of-arrival sensors coupled with IMU sensors, head and gaze tracking, open up interesting possibilities for capturing both point-of-view visual information and the individual’s physiological and neurological states. Together with their low power requirements and connectivity capabilities, such devices can provide cues about individual's intents and states, at real-time, to their robotic teammates for implicit coordination in various scenarios including battlefields and Industry 4.0 settings. 

\item \emph{\textbf{Highly Mobile, SWAP-Constrained Unmanned Vehicles: }} Robotic platforms that are size, weight, and power constrained are attractive for human-machine teams that operate side by side, occupying the same physical space. While high levels of mobility can pose serious safety concerns for the human agents, the physical configurations of SWAP-constrained platforms make them safer options without compromising on the sensing capabilities they can offer. However, such platforms can be extremely restrictive in terms of performing fully autonomous navigation, sensing and onboard computation for sense-making. We posit that such platforms can leverage guidance from their human partners for intelligently adapting between \emph{autonomous} and \emph{assisted} operations for longer operation windows without compromising on the sensing efficacy.
\end{itemize}

% In this work, we make the following contributions:
% \begin{enumerate}
%     \item Propose and ideate scenarios where wearable-based, continous monitoring of human cognitive states can be exploited for machine adaptation in mixed human--machine teams.
%     \item Identify key challenges and open questions in realizing continuous human-machine coordination and adaptation for autonomous teams, and provide initial solutions.
%     \item Demonstrate feasibility of such machine adaptation using benchmark wearable, physiological sensing datasets.
%     \item Discuss open questions, limitations of the current state-of-the-art and provide a technical road map.
% \end{enumerate}
\subsection{Motivating Scenarios}
``Passive guidance" from human(s) in the team can help direct machines for maximizing the effectiveness of human-machine combined sensing objectives. Communication, the process of information exchange, between man and machine is key for successful team performance~\cite{stowers2021improving}. While advances in natural language processing or exchange of visuals aid in more direct, explicit communication between teammates, "implicit coordination" where machines are able to synchronize with their human-teammates without explicit intervention has its advantages. Previous studies~\cite{macmillan2004human} show that implicit coordination is helpful under high workload situations due to a reduction in communication overheads and the resulting distractions. Here, we describe two scenarios where we envision tightly-coupled and responsive drones that adapt to human intent based on implicit guidance to be highly effective. We illustrate this in Figure~\ref{fig:tagteam-modes}.

\textbf{Attention-Shared, coordinated visual scanning for reconnaissance and search-and-rescue missions: }
An exemplar scenario is where a dismount is teamed up with a robotic teammate, with the robotic teammate equipped with a variety of sensors such as RGB cameras and LIDAR. Coordinated scanning can include two goals: (1) complementary - achieve wider coverage where the machine is able to scan regions where the human is not paying attention to, and as a team, achieve faster and efficient scanning, and (2) collaborative supplementary – where the machine provides enhanced resolution scanning and inference when a human requires more accurate visibility and augmentation from the machine. In both cases, machines require to know where the human is paying attention to, and not just what is within the "visible" range. A collection of wearable sensors can help intelligently infer the type of assistance the human teammate requires (e.g., whether complementary or supplementary) as well as in passively guiding the drone to areas that need augmented attention and/or at varied perception configurations (e.g., resolution, coverage, etc.).

\textbf{Non-verbal, interactive communication for continuous learning}
While natural language is a more direct interface for communication with machines, non-language behaviours (voice quality, body language, etc.)~\cite{duncan1969nonverbal} and motor correlates of speech and verbal communication (e.g., gaze, facial expressions, gestures~\cite{mavridis2015review} play a crucial role in effective communication of the human expressor. The ability to sense such cues can be beneficial in many battlefield scenarios including (1) machine learning and adaptation with passive human reinforcement (e.g., via affirmative or negative thinking inferred from brain activity~\cite{ruf2013semantic}) and (2) to interactively resolve comprehension ambiguities of human-to-robot instructions (e.g., gestures to zoom in/out for a vision sensing task, corrective or altered behaviours based on cues such as a frown or shaking of the head).
% \section{Key Technical Challenges and Opportunities}
\\
\textbf{Key Contributions: }Through this initial work,  we articulate a paradigm of highly interactive, attention-shared human-drone teams and identify key technical capabilities such teams require to address. We also share details of an initial prototype we built and share early results from enabling accurate motion transfer from human to drone.

\section{Design Goals}
\label{sec:sysoverview}
To support the \name scenarios that we envision, we enumerate the following key design goals that human--drone teaming systems should achieve.

\subsection{Wearable-based attentive state estimation: }
\textbf{Visual attention: }In this work, we will develop techniques to gauge human visual attention using a combination of wearable technologies that allow for accurate tracking of eye movements (e.g., using noisy EOG signals) in the presence of motion artifacts (determined using on-body inertial and EEG signals) and investigate techniques for continuous, light-weight exchange of attention information. The machines then adapt their attentional focus and/or resolution, on-the-fly, to synchronize with their human-teammates’ intent. Evaluations of such systems will require both recreations of dynamic environments in an augmented-reality based experimental setups and real-world studies to study the trade-offs between accuracy on vision tasks, energy efficiency and latency, baselining against attention-agnostic models for both the complementary and collaborative goals. We believe that these models of attention-responsive adaptation of machine intelligence will not only improve the overall inference and situational intelligence accuracy, but also provide a practical way to reduce energy and computational requirements, thereby enabling longer operational lifetimes and more ergonomic machine form factors.

\textbf{Non-verbal intent: }A key technical capability that this proposed work builds is how such nonverbal intent can be reliably measured with a combination of wearable sensors including EEG sensors, on-body inertial sensors, gaze trackers, etc. While the effectiveness of physiological sensing has been demonstrated with high-fidelity sensors, in lab and controlled settings, for tasks such as inferring emotion, the ability to infer nonverbal cues using energy-efficient, but sparse, physiological signals is relatively under-explored (e.g., a 2-channel Emotiv MN8 earable as opposed to a 64-channel Biosemi ActiveTwo EEG sensor). More recently, ear-worn inertial sensors have indeed been shown to be effective in detecting activities such as head and neck movements~\cite{hossain2019human} and more finer-grained motion such as tapping and sliding of the teeth~\cite{prakash2020earsense}. Furthermore, recent work~\cite{weerakoon2022cosm2ic} has demonstrated that pointing-gesture based input in combination with visual and verbal inputs can improve accuracy of object picking tasks. To this end, the work will explore sensor fusion of wearable modalities (e.g., physiological, micro-expressions, lip/jaw movements based on inertial measurements, gaze tracking with smart glasses, etc.) for accurate nonverbal cue sensing and its application in the two example scenarios. One such possible use case is an urban battlefield where the human dismount issues commands to a robotic assistant, for example, to investigate some roadside objects for possible threats. While ongoing work explores such multi-modal instruction comprehension for civilian environments (e.g., factory floors), kinetic military environments are likely to be characterized by higher levels of stress, distraction and time constraints. This in turn will affect the ways in which humans communicate instructions and raise the importance of factoring in non-verbal input (e.g., stress or fear levels) in defining the performance requirement (not just accuracy but factors such as time sensitivity) for such comprehension. We will thus engineer a suite of multimodal features for light-weight detection of higher order non-verbal cues, develop an adaptive system that orchestrates the triggering of multiple modalities based on energy, accuracy and task-specific performance (e.g., how fast can the human and machine converge on instruction comprehension?) trade offs, baselining against purely language-based comprehension.  

\subsection{Global understanding of location and motion: }For highly mobile situations such as scouting, search-and-rescue, the human and agent require highly accurate spatial and motion awareness. This a key requirement, especially in previously unseen environments such as those envisioned. In our scenarios, we require that the drone or agent be capable of mimicking, or closely follow, the human for situations such as providing blindspot coverage. In situations where the drone detects that the human requires complementary or supplementary scanning, through the combination of various wearable-based attentive states, the drone navigates independent of the human for completing such tasks, and \emph{boomerangs} back to its human accomplice upon completion of the task. Such coordination requires a common grounding of the spatial coordinate systems of the two devices (human-worn wearable and the drone) and continuous tracking.

\subsection{Constraint-aware orchestration: }Cooperating human-machine teams can be high efficient in terms of scanning target areas in shorter windows, as opposed to a human-alone or drone-alone baseline. Whilst, small unmanned vehicles, especially aerial vehicles, are extremely restrictive in terms of their size, weight, and hence power and memory, wearable devices such as the Hololens 2 are much more resilient (lasts up to 2-3 hours with a single charge and active rendering throughout the duration).

\section{Initial Prototype}
\label{sec:imp}
\begin{figure*}[t]
\centering
        
	  \centering \includegraphics[scale=0.05]{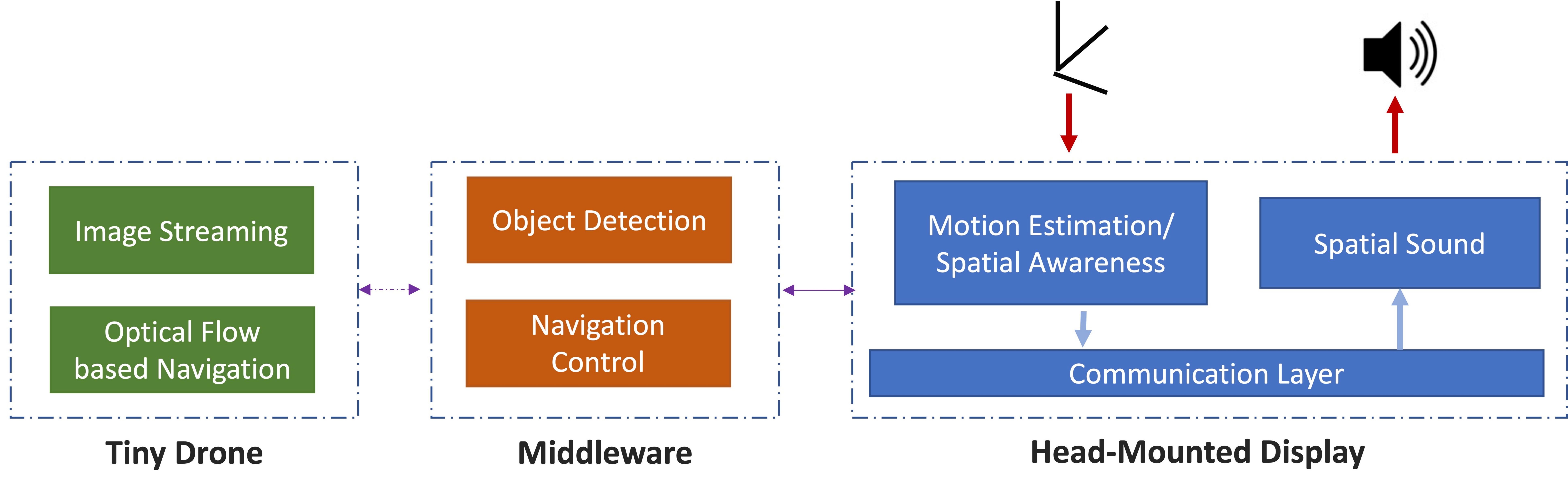}
      \caption{\name Components.} 
      %\vspace{-0.2in}
      \label{fig:sysoverview}
\end{figure*}

\begin{figure}[t]
\centering
        
	  \centering \includegraphics[scale=0.04]{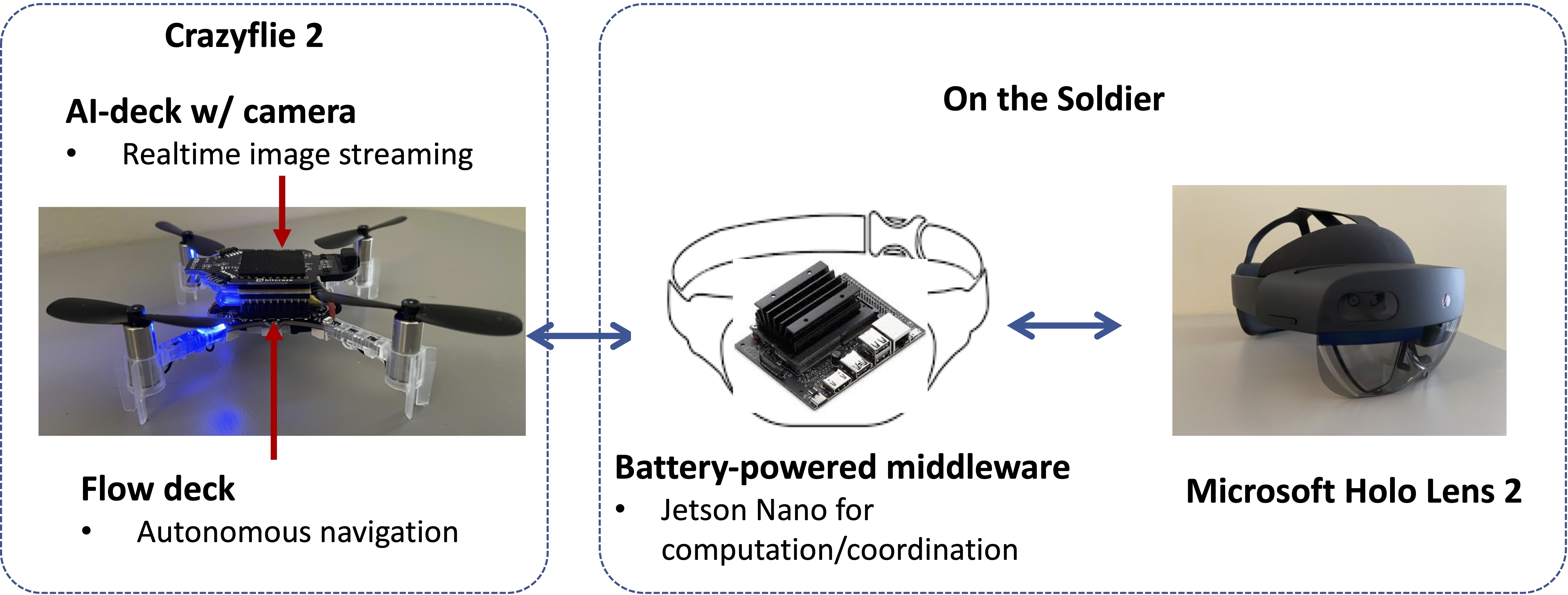}
      \caption{\name Implementation.} 
      %\vspace{-0.2in}
      \label{fig:implementation}
\end{figure}

Towards realizing our vision for highly interactive human--drone teams, we describe our efforts in prototyping an early version of \name that accomplishes real-time motion awareness for the default scenario of \textit{blindspot detection}. We present a simplified architecture in Figure~\ref{fig:sysoverview}. The system consists of the following components (see Figure~\ref{fig:implementation}).

\begin{figure}[t]
\centering
        
	  \centering \includegraphics[scale=0.035]{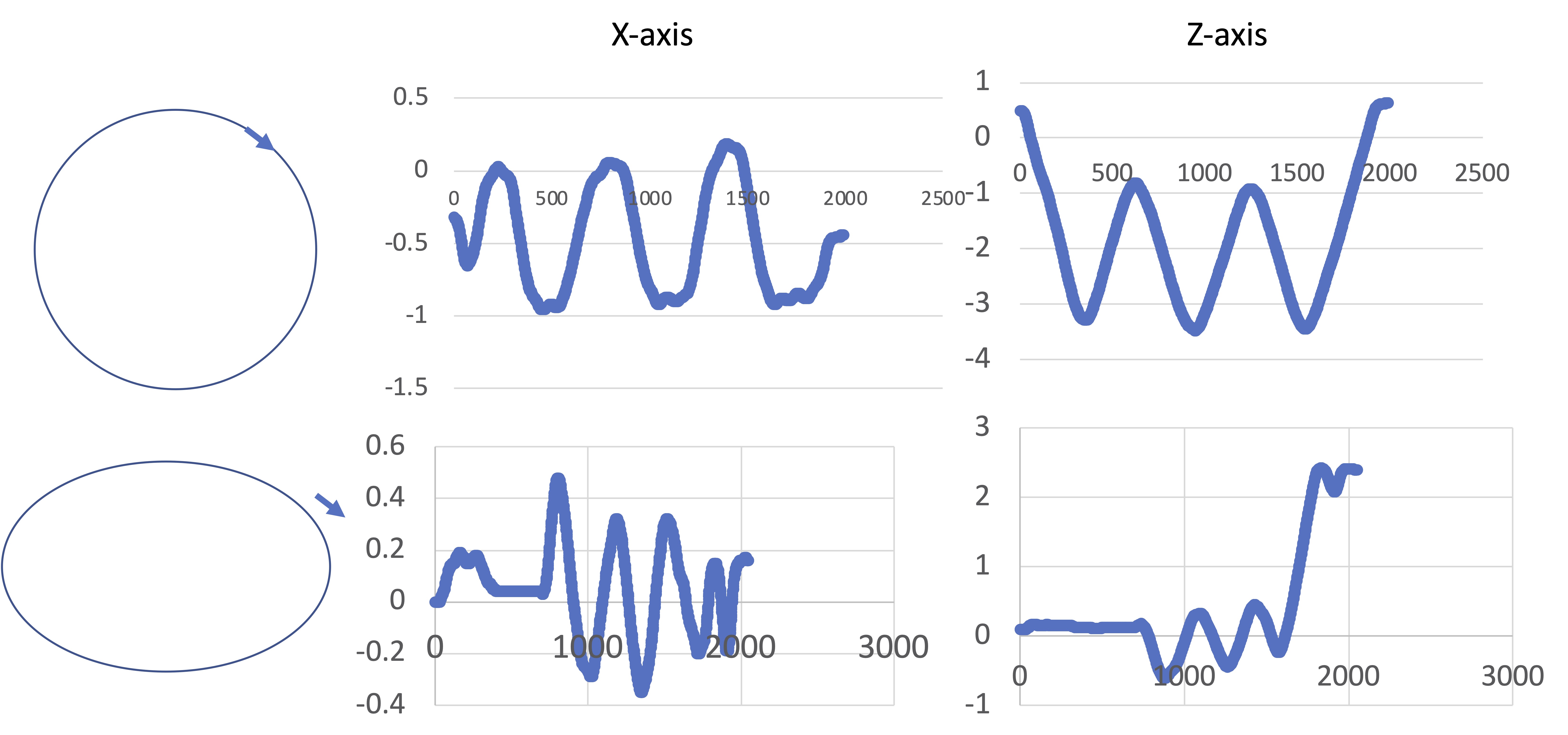}
      \caption{Coordinate variations for exemplar motions: circular (top) and oval (bottom)} 
      %\vspace{-0.2in}
      \label{fig:hl-motion}
\end{figure}

\textbf{A tiny drone: }The Crazyflie 2.1~\cite{crazyflie}, a nano class quadcopter, is an open source flying development platform. It is an example of a SWAP-constrained edge device similar to what we envision in our scenarios, that fits inside the palm of the hand and weighs only a nominal 27 grams. The base design of the quadcopter consists of a STM32F405 main application MCU (microcontroller unit) and communication enabled via a nRF51822 radio and power management MCU. With the  long range open USB radio dongle based on the nRF24LU1+ from Nordic Semiconductor~\footnote{https://www.bitcraze.io/products/crazyradio-pa/} (supporting 2.4GHz ISM band and Bluetooth Low Energy), the drone can be controlled from distances as far as 1 km, under ideal conditions. The capabilities of the Crazyflie 2.1 can be extended using a range of off-the-shelf or custom-built expansion decks. In this work, we use the AI deck~\footnote{https://www.bitcraze.io/products/ai-deck/} for capturing and wirelessly streaming images and the flow deck V2~\footnote{https://www.bitcraze.io/products/flow-deck-v2/} for stable flight.

\textbf{Head-mounted Wearable Display (Smart helmet): }We use the Microsoft HoloLens 2, a sensor-rich smart wearable, in our implementation. Through its four grayscale head-tracking cameras, the depth camera and a combination of IMU sensors, the HoloLens device has the capability to maintain highly accurate spatial and motion awareness of the wearer. In the current prototype, we use the $\text{Camera.main.transform}$ property of the main camera to localize the user; this property initializes to (0, 0, 0) at the real-world position when the App is launched. In Figure~\ref{fig:hl-motion}, we plot the variation along the $x$ and $z$ directions for two exemplar, controlled motions, circular and oval, performed by a human subject.   

\textbf{Middleware: }The Nvidia Jetson Nano\footnote{https://developer.nvidia.com/embedded/jetson-nano-developer-kit} acts as the middleware between the drone and the wearable on the soldier. The Nano is a mobile device consisting of a 128-core Maxwell GPU and a Quad-core ARM A57 CPU (clock speed of 1.43 GHz) and 4 GB system memory. In the current implementation, the middleware acts as a broker between the two devices; (a) it ascertains positional information from the wearable device over the MQTT~\cite{hunkeler2008mqtt} protocol, by subscribing to a certain topic, (b) transforms locational information from the HoloLen's coordinate system to the Crazyflie 2.1's coordinate system, and (c) performs object detection (using the SSD MobileNet V2 detector~\cite{sandler2018mobilenetv2}) on the stream of images transferred by the drone to provide real-time indication of objects in the human's blindspots. Such detections are published as MQTT back to the HoloLens which outputs the information as \emph{Spatial Sound}~\footnote{https://docs.microsoft.com/en-us/windows/mixed-reality/design/spatial-sound}, dependent on the distance and angle of the object relative to the human. During the transformation stage (Step (b)), we take a dead-reckoning based approach~\cite{steinhoff2010dead} where we assume that the initial positions of the HoloLens 2 (or the human) and the drone are known, and we estimate the change in movement needed by the drone, from the change in movement sensed by the HoloLens, assessed periodically, and task the drone to \emph{move} to the new position with a certain velocity (calculated as the distance moved divided the time period between updates). Through experimentation, we found the appropriate coordinate transformation between the two devices to be: $X_{hololens} \rightarrow Z_{crazyflie}$ and $Z_{hololens} \rightarrow -X_{crazyflie}$ (note that in our implementation, the Crazyflie drone faces backwards to provide situation awareness in the human's blindspot regions).

\section{Preliminary Results}
\label{sec:eval}
\begin{figure}[t]
\centering
        
	  \centering \includegraphics[width=0.6\linewidth]{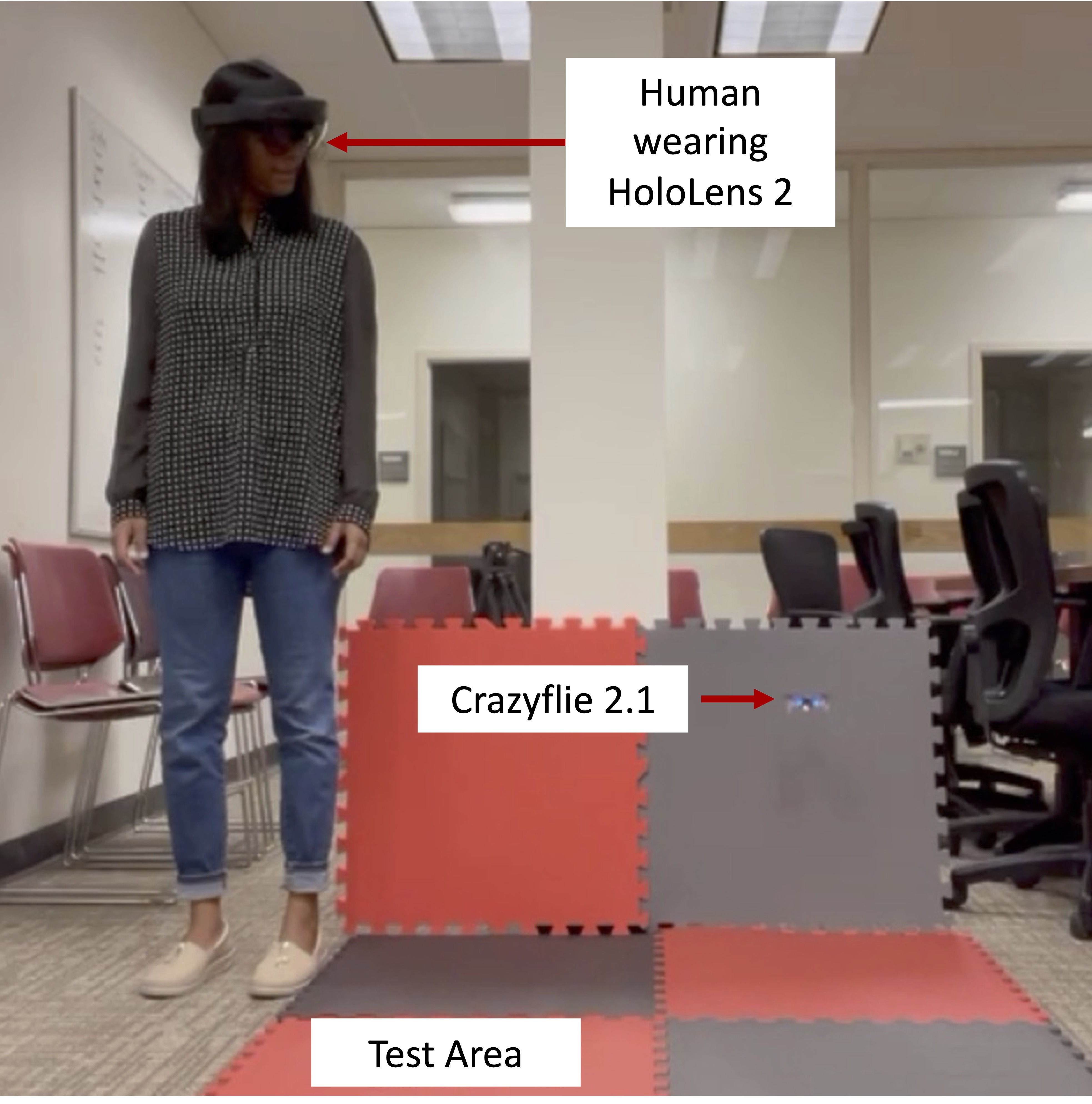}
      \caption{Test setup for location tracking. See \hl{\hbox{https://youtu.be/INQULM5csMU}} for a video demonstration. }
      %\vspace{-0.2in}
      \label{fig:testarea}
\end{figure}

Figure~\ref{fig:testarea} shows the setup (a 4 ft $\times$ 6 ft area) we use in establishing the feasibility of tracking (the human's) movements using the dead-reckoning based approach. As seen in the video demonstration, the drone is able to accurately mimic the movements of the human. 

To quantify this, we use the vatic.js tool~\footnote{https://stefanopini.github.io/vatic.js} for annotating the bounding box representing both the human's head as well as the drone. In total, the video consists of  503 frames. We use Dynamic Time Warping to capture the similarity in the trajectories of the head and the drone, and observe a high level of synchronization ($\approx$ 0.92 where 1 is the highest).  

\section{Discussion and Open Problems}
In our current implementation we tackle a fundamental and necessary building block towards building \name (i.e., motion awareness between human and drone). As we describe earlier in Section~\ref{sec:sysoverview}, the paradigm of attention-sharing, human-drone teams require several other technical capabilities such as inferring human intent and attention automatically, implicit coordination between the human and drone, and orchestration of the collective resources (e.g., on-board computation on the devices, network bottlenecks, etc.) for effective cooperation and communication. Beyond these immediate requirements, we enumerate a number of open problems for advancing research in this direction.

\textbf{Extending to multi-human, multi-drone \emph{teams}: }While our current scenarios consider a pair of agents, human and otherwise, many of these generalize to multi-human, multi-drone teams -- for instance, a SWAT team involved in a high-risk search cooperating with a swarm of drones. Such extensions impose additional challenges such as multi-source information aggregation and dissemination under potentially conflicting sources, data prioritization for processing and exchange under scheduling constraints, and physical-formation aware coordination. 

\textbf{Aspects of trust and psychology: }While the form factors of sensing and processing platforms have evolved for \names-like technical capabilities to be possible. Whilst works such as those of Hancock et al.~\cite{hancock2011meta} study aspects of trust and psychology for human-robot interaction scenarios in general, the human factors related to close-contact human-drone teams remain largely under studied.
\section{Related Work}
\textbf{Human--Drone Interaction (HDI): }Since recently, researchers have started looking at the technical capabilities HDI requires and novel use cases that it enables~\cite{christ2016human, pfeiffer2022visual}. In early works, Christ et al.~\cite{christ2016human} study the impact of the level of autonomy of the drone on user experience. PFeiffer et al.~\cite{pfeiffer2022visual} demonstrate that understanding where humans pay visual attention can aid in improving the navigation capabilities of drones through imitation learning. Several works have specifically focused on exploring \emph{modalities} for interaction such as gesture-based control (e.g., using smart gloves~\cite{muezzinouglu2021intelligent}, ~\cite{cauchard2015drone}), voice~\cite{fayjie2017voice, Menshchikov, landau2017system}, gaze-based teleoperation~\cite{hansen2014use, yu2014human}, etc. While these works explore the \emph{explicit control} of the drone through these modalities, in our work, we emphasize the need for \emph{implicit} guidance where the drone automatically infers the intent of the human -- for instance, using gaze-based features such as saccades and dwell time to infer areas of uncertainty where the drone should perform a secondary scan to maintain higher accuracy of the scan. Our work is the first, to the best of our knowledge, to articulate a paradigm of humans and drones working in close proxemics to perform cooperative tasks such as indoor scouting and search-and-rescue, whilst addressing challenges in terms of enabling real-time behavior adaptation of the drone leveraging on multi-modal, yet implicit interactions between the drone and its human partner.

\textbf{Collaborative sense-making at the edge: }There have been a number of efforts in enabling lightweight collaboration between machines, especially for compute-intensive tasks such as machine perception using deep neural networks (DNNs). Several recent works have explored optimization techniques for networked sensors to achieve efficient querying~\cite{hung2018videoedge, jiang2018chameleon}. Recent works~\cite{qiu2018kestrel, lee2018, hotmobile2019, eugene2019} have also explored the idea of selective activation of nodes in a group of collaborating sensors
– e.g., Qiu et al.~\cite{qiu2018kestrel} describe a vehicle tracking scenario
where mobile nodes in a hybrid (mobile/infrastructure) camera
network are activated selectively, only to resolve ambiguities.
Jain et al.~\cite{hotmobile2019} provide preliminary examples of the possibility
of using inputs from peer, overlapping cameras to utilize such
spatiotemporal correlations to optimize the video analytics
pipeline. The idea of collaboration among AIoT devices at
the edge, and its attendant challenges, has also been mooted
more generally recently~\cite{eugene2019, hannaneh2019}. Most recently, ComAI~\cite{comai2022} demonstrates concrete
mechanisms for low-overhead collaboration for perception. To the best of knowledge, this work is among the first to prescribe the need for lightweight collaboration for highly effective human--machine mixed teams.

\section*{Acknowledgement}
We acknowledge the support of the U.S. Army Grant No. W911NF21-
20076.

% \small
\bibliographystyle{abbrv}
\bibliography{main} 

\end{document}